\documentclass[prd,preprintnumbers,amsmath,amssymb]{revtex4}

\input epsf
\begin{document}
\draft

\title{Discussion on Event Horizon and Quantum Ergosphere of Evaporating Black Holes in a Tunnelling Framework}
\author{Jingyi Zhang} \affiliation{ Center for Astrophysics,
Guangzhou University, 510006, Guangzhou, China (E-mail:
physicz@yahoo.cn)}
\author{Zheng Zhao} \affiliation{ Department of physics,
Beijing Normal University, 100875, Beijing, China (E-mail:
zhaoz43@hotmail.com)}
\date{\today}

\begin{abstract}
In this paper, with the Parikh-Wilczek tunnelling framework the
positions of the event horizon of the Vaidya black hole and the
Vaidya-Bonner black hole are calculated respectively. We find that
the event horizon and the apparent horizon of these two black
holes correspond respectively to the two turning points of the
Hawking radiation tunnelling barrier. That is, the quantum
ergosphere coincides with the tunnelling barrier. Our calculation
also implies that the Hawking radiation comes from the apparent
horizon.
\end{abstract}
\maketitle
\vskip2pc

\section{introduction}
In 1974, Hawking published his astounding discovery that black
holes radiate like a black body\cite{Hawking1,Hawking2}. Since
then, black hole thermodynamics become a very interesting topic in
theoretical physics. Since black hole radiates thermally, it must
have entropy and other thermal properties. According to Hawking's
original paper, the radiation comes from the vacuum fluctuation
near the horizon. The idea is that when a virtual particle pair is
created just outside the horizon, the negative energy virtual
particle, which is forbidden outside, can tunnel inwards, and the
positive energy virtual particle materializes as a real particle
and escape to infinity. However, a black hole usually has event
horizon and apparent horizon. Where does the Hawking radiation
come from? At present, there are two different viewpoints about
this problem. One is that the event horizon constitutes a black
hole, and Hawking radiation comes from this surface or the
neighborhood\cite{Hawking3,York}. In
Ref.\cite{Zhao1,Zhao2,Zhao3,Zhao4,Zhang0} an expression of local
event horizon was obtained by assuming that the local event
horizon is a null hypersurface and preserves the symmetries of the
space time. Another viewpoint is that Hawking radiation comes from
the apparent horizon, the laws of black hole thermodynamics and
the expression of the Bekenstein-Hawking entropy are only suitable
to this surface
\cite{Hajicek,Hayward,Cai,Liu01,Liu02,Liu03,Liu04}. For a
stationary black hole, the event horizon and the apparent horizon
coincide with each other, these two different viewpoints do not
cause problem. However, for a non-stationary black hole, the event
horizon and the apparent horizon deviate from each other.
Therefore, for non-stationary black holes it is very important for
us to determine where the Hawking radiation comes from. In this
paper we attempt to investigate this problem with Parikh-Wilczek
tunneling framework and calculate the concrete positions of these
two horizons in the Vaidya black holes and Vaidya-Bonner black
holes. In General Relativity, the non-stationary black holes have
their static counterparts, such as the Vaidya black holes
correspond to the Schwarschild black hole, and the Vaidya-Bonner
black holes to the Reissner-Nordstr\"{o}m black holes. In
Parikh-Wilczek tunnelling framework\cite{Parikh1,Parikh2,Parikh3},
the static background space time, such as the Schwarzschild black
hole, is dynamical when self-gravitation of tunnelling particles
is taken into account. Therefore, the treatment about the quantum
tunnelling in Parikh-Wilczek's tunneling framwork describes the
evolution of the non-stationary counterpart black hole. In this
paper, by treating the tunnelling process of the Schwarzschild
black hole as a volution of the Vaidya black hole, we first
calculate the position of the event horizon of the Vaidya black
holes, and then, we discuss the ergosphere and the location where
Hawking radiation taking place. Similarly, the event horizon of
the Vaidya-Bonner black holes is discussed too.
\section{event horizon and quantum ergosphere of the Vaidya black holes}

Vaidya black hole is the simplest non-stationary black hole. Its
line element is as follows
\begin{equation}
ds^2=-(1-\frac{2M(v)}{r})dv^2+2dvdr+r^2d \Omega ^2.\label{line1}
\end{equation}
Ref.\cite{Zhao1,Zhao2,Zhao3,Zhao4,Zhang0} give an equation about
the local event horizon by suggesting that the event horizon is a
null supersurface and preserves the symmetry of the space time,
namely,
\begin{equation}
r_{H}=\frac{2M(v)}{1-2\dot{r}_{H}},\label{horizon1}
\end{equation}
where
\begin{equation}
\dot{r}_{H}=\frac{dr_H}{dv}.
\end{equation}
The apparent horizon $r_{AH}$ and the timelike limit surface
$r_{TLS}$ can be easily obtained respectively from the equations
\begin{equation}
\Theta|_{r=r_{AH}}=0,
\end{equation}
and
\begin{equation}
g_{vv}=0,
\end{equation}
where $\Theta$ is the expansion of the null rays. Namely,
\begin{equation}
r_{AH}=r_{TLS}=2M(v).\label{AH}
\end{equation}
Obviously, Eq. (\ref{horizon1}) is a to be solved equation about
the event horizon. Generally, it is very difficult for us to
obtain the exact expression of the event horizon. In the following
we will try to obtain the concrete position by using the
Parikh-Wilczek tunnelling framework.

In 2000, Parikh and Wilczek pointed out that Hawking's previous
calculation about the black hole radiation have a defect: energy
conservation was not enforced during the emission process of a
tunnelling particle\cite{Parikh1,Parikh2,Parikh3}. If we consider
the energy conservation, the particle's self-gravitation should be
taken into account. Then, the background space time should be
dynamical, and therefore the spectrum of the Hawking radiation
will not be
thermal\cite{Parikh1,Parikh2,Parikh3,Zhang3,Zhang4,Hemming,Medved,Alves,Vagenas1,Vagenas2,Vagenas3,Vagenas4,Vagenas5,Vagenas6,Vagenas8,Zhang1,Zhang2,Liu,Wu,Zhang5,Zhang6,Zhang7,Zhang8,Zhang9,Zhang10,Zhang11,Zhang12,Zhang13,Zhou,Liu2,Hu1,Hu2,He,Jiang,Majhi1,Majhi2,Majhi3,Kar}.
That is, when a particle (in order to preserve the symmetry of the
space time, for the sake of simplicity, we take the tunneling
particle as a sphere shell,a s-shell, as did in
Ref.\cite{Parikh1,Parikh2,Parikh3}) with mass $\omega$ tunnels
out, the effect background space time will become a Schwarzschild
black hole with the mass $M-\omega$. Therefore, the background
space time is in fact a evaporating Vaidya black hole space time.
We can calculate the position of the event horizon of the Vaidya
black hole by using the Parikh-Wilczek tunnelling framework.

let us first take into account a Schwarzschild black hole with the
mass M. Its line element is as follows
\begin{equation}
ds^2=-(1-\frac{2M}{r})dt^2_s+(1-\frac{2M}{r})^{-1}dr^2+r^2d \Omega
^2.\label{line2}
\end{equation}
In the Parikh-Wilczek tunnelling framework, in order to calculate
the emission rate of the tunnelling particles, Painlev\'{e}
coordinates are adopted, that is, the line element of the
Schwarzschild black hole space time should be written as the
following form
\begin{equation}
ds^2=-(1-\frac{2M}{r})dt^2+2\sqrt{\frac{2M}{r}}dtdr+dr^2+r^2d
\Omega ^2.\label{line3}
\end{equation}
However, to compare with the Vaidya black hole in literature, in
this paper we adopt the ingoing Eddington-Finkelstein coordinates
($v,r,\theta,\varphi$) by the transformation
$v=t+r+\frac{1}{2M}\ln{(\frac{r}{2M}-1)}$ from the Schwarzschild
coordinate system $(t_s,r,\theta,\varphi)$\cite{Ren}. In this
coordinate system, the line element reads
\begin{equation}
ds^2=-(1-\frac{2M}{r})dv^2+2dvdr+r^2d \Omega ^2.\label{line4}
\end{equation}
Obviously, if the mass of the black hole is a function of $v$,
that is $M=M(v)$, then, the Eq.(\ref{line4}), the line element of
the Schwarzschild black hole, will be the same as that of the
Vaidya black hole. Similar to the Painlev\'{e} coordinate system,
the ingoing Eddington-Finkelstein coordinate system have a series
of good properties, and we can use it to describe the tunnelling
process and calculate the emission rates of the tunnelling
particles\cite{Ren}.

As mentioned above, in Hawking's original paper, Hawking radiation
is described as a tunneling process triggered by vacuum
fluctuations near the horizon. The idea is that when a virtual
particle pair is created just outside the horizon, the negative
energy virtual particle, which is forbidden outside, can tunnel
inwards, and the positive energy virtual particle materializes as
a real particle and escape to infinity. This process of tunneling
can also be equivalently explained by Paikh and Wilczek as the
image: there is a virtual particle pair created just inside the
horizon, the positive energy virtual particle can tunnel out-no
classical escape rout exists-where it materializes as a real
particle. In either case, the negative energy particle is absorbed
by the black hole, resulting in a decrease in the mass of the
black hole, while the positive energy particle escape to infinity,
appearing as Hawking radiation. According to Parikh-Wilczek
tunneling framework, when a particle with a mass of $\omega$
tunnels out, the black hole will shrink from $r_H(M)$ to
$r_H(M-\omega)$. It is the contraction of the black hole or the
self-gravitation of the tunnelling particle set the barrier. In
this model, the positions $r_H(M)$ and $r_H(M-\omega)$ correspond
to two turning points. If we treat the dynamical Schwarzschild
black hole as an evaporating Vaidya black hole, then, we can
calculate the position of the event horizon.

According to the Parikh-Wilczek tunnelling framework, when a
particle (a shell) tunnels out, the mass of the black hole
decreases. Then, the event horizon will shrink. The tunnelling and
the shrinking take place at the same time. Naturally, the
tunnelling speed of a particle (a s-shell) is equal to that of the
shrinking of the black hole. Therefore, we can obtain the
shrinking velocity of the event horizon, namely,
\begin{equation}
\dot{r}_H=-\dot{r},\label{r1}
\end{equation}
where $\dot{r}$ denote the velocity of the tunnelling particle
(s-shell). In fact, for a s-shell, or a s wave, it is the velocity
of the wave front.

There are two types of tunnelling particles. One is the massless
particles, or null particles, which travel along a null geodesic.
For this type of particles, the velocity of the wave front can be
obtained by letting $ds^2=d\theta=d\varphi=0$, namely \cite{Parikh1,Parikh2,Parikh3},
\begin{equation}
\dot{r}=\frac{1}{2}(1-\frac{2M}{r}).\label{r2}
\end{equation}
Another type of tunnelling particles are the non-zero-rest mass
particles, or massive particles. Dislike the null particles, the
tunnelling of the massive particles do not travel along the null
geodesics. We can treat a massive particle as a de Broglie wave,
and in order to preserve the spherical symmetry of the space time
during the tunnelling process, we treat it as a spherical wave.
Like the treatment in Ref.\cite{Zhang3}, the velocity of the wave
front of the de Broglie s-wave is
\begin{equation}
\dot{r}=-\frac{1}{2}\frac{g_{00}}{g_{01}}=\frac{1}{2}(1-\frac{2M}{r}).\label{r3}
\end{equation}
Eqs. (\ref{r2}) and (\ref{r3}) show that the velocities of the
wave fronts for the massless or massive particles are the same.
Considering the self-gravitation, the mass $M$ in Eqs.
(\ref{line4}),(\ref{r2}),and (\ref{r3}) should be replaced with
$M-\omega$. Therefore, we have
\begin{equation}
\dot{r}_H=-\dot{r}=-\frac{1}{2}(1-\frac{2(M-\omega)}{r}).\label{r4}
\end{equation}
In the following, we will calculate the concrete position of the
event horizon of the Vaidya black hole by using Eq. (\ref{r4}). As
described above, we take the dynamical Schwarzschild black hole as
a Vaidya black hole. Before the particle tunnels out, the space
time is a stationary Schwarzschild black hole space time, the
event horizon, the apparent horizon, and the timelike limit
surface locate at the same place, that is,
\begin{equation}
r_H=r_{AH}=r_{TLS}=2M.\label{r5}
\end{equation}
When particles tunnel out, the space time will become dynamic if
self-gravitation is taken into account. To calculate the shrinking
velocity of the event horizon at $r=r_{AH}$, we take $r=2M$. Then,
the shrinking velocity of the event horizon is
\begin{equation}
\dot{r}_H=-\dot{r}|_{2M}=-\frac{\omega}{2M}.\label{r6}
\end{equation}
Therefore, the position of the event horizon of the Vaidya black
holes is
\begin{equation}
r_{H}=\frac{2M(v)}{1-2\dot{r}_{H}}=\frac{2M}{1+\frac{\omega}{M}}\approx
2(M-\omega).\label{horizon2}
\end{equation}
Obviously, the event horizon $r_H$ is a turning point of the
tunnelling barrier. Since the apparent horizon and the timelike
limit surface of the Vaidya black hole locate at the same place,
and correspond to another turning point, namely,
\begin{equation}
r_{AH}=r_{TLS}=2M,\label{r7}
\end{equation}
we see that the ergosphere region coincide with the tunnelling
barrier. That is, the ergospere region corresponds to the
classical inhibited region.
\section{event horizon and quantum ergosphere of the Vaidya-Bonner black holes}
The line element of the Vaidya-Bonner black hole is
\begin{equation}
ds^2=-(1-\frac{2M(v)}{r}+\frac{Q^2(v)}{r^2})dv^2+2dvdr+r^2d \Omega
^2.\label{line7}
\end{equation}
Ref.\cite{Zhu} gives the expression of the local event horizon by
suggesting that the event horizon is a null hypersurface and
preserves the symmetry of the space time, namely,
\begin{equation}
r_{H}=\frac{M+\sqrt{M^2-Q^2(1-2\dot{r}_H)}}{1-2\dot{r}_{H}},\label{horizon3}
\end{equation}
where
\begin{equation}
\dot{r}_{H}=\frac{dr_H}{dv}.
\end{equation}
The apparent horizon and the timelike limit surface locate at
\begin{equation}
r_{AH}=r_{TLS}=M+\sqrt{M^2-Q^2}.\label{AH2}
\end{equation}
Let us consider the corresponding stationary balck hole, the
Reissner-Nordstr\"{o}m black hole. Its line element is
\begin{equation}
ds^2=-(1-\frac{2M}{r}+\frac{Q^2}{r^2})dt^2+(1-\frac{2M}{r}+\frac{Q^2}{r^2})^{-1}dr^2+r^2d
\Omega ^2.\label{line22}
\end{equation}
There are two event horizons, and their expressions are
\begin{equation}
r_{\pm}=M\pm\sqrt{M^2-Q^2}.\label{AH22}
\end{equation}
We can easily get the surface gravity on the event horizon,
\begin{equation}
\kappa_{\pm}=\frac{r_+-r_-}{2r_{\pm}^2}.\label{k}
\end{equation}
In order to compare with the Vaidya-Bonner black hole, we adopt
the ingoing Eddington-Finkelstein coordinate system
($v,r,\theta,\varphi$). Let us do the transformation $v=t+r_*$,
where
\begin{equation}
r_*=r+\frac{1}{2\kappa_+}\ln{\frac{r-r_+}{r_+}}-\frac{1}{2\kappa_-}\ln{\frac{r-r_-}{r_-}}.\label{rr}
\end{equation}
Then, the line element of the Reissner-Nordstr\"{o}m black hole
can be written as
\begin{equation}
ds^2=-(1-\frac{2M}{r}+\frac{Q^2}{r^2})dv^2+2dvdr+r^2d \Omega
^2.\label{line8}
\end{equation}
Comparing Eq. (\ref{line8}) with Eq. (\ref{line7}) we see that the
Reissner-Nordstr\"{o}m is a static counterpart of the
Vaidya-Bonner black hole. Similarly, there are two types of
tunnelling particles, the massless s-shell and the massive
s-shell. The velocity of the wave front of the massless shell can
be obtained by letting $ds^2=d\theta=d\varphi=0$, namely,
\begin{equation}
\dot{r}=\frac{1}{2}(1-\frac{2M}{r}+\frac{Q^2}{r^2}).\label{r22}
\end{equation}
For the massive particles, like the treatment in
Ref.\cite{Zhang3}, the velocity of the wave front of the de
Broglie s-wave is
\begin{equation}
\dot{r}=-\frac{1}{2}\frac{g_{00}}{g_{01}}=\frac{1}{2}(1-\frac{2M}{r}+\frac{Q^2}{r^2}).\label{r23}
\end{equation}
Eqs. (\ref{r22}) and (\ref{r23}) show that the velocities of the
wave front for the massless or massive particles are the same.
Considering the self-gravitation, the mass $M$ in Eqs.
(\ref{line8}),(\ref{r22}),and (\ref{r23}) should be replaced with
$M-\omega$. Therefore, we have
\begin{equation}
\dot{r}_H=-\dot{r}=-\frac{1}{2}(1-\frac{2(M-\omega)}{r}+\frac{Q^2}{r^2}).\label{r24}
\end{equation}

Let $r=r_{AH}=M+\sqrt{M^2-Q^2}$, then, the shrinking velocity of
the event horizon can be written as
\begin{equation}
\dot{r}_H=-\dot{r}|_{r_{AH}}=-\frac{\omega}{M+\sqrt{M^2-Q^2}}.\label{r6}
\end{equation}
Substituting Eq. (\ref{r6}) into Eq. (\ref{horizon3}), we get the
position of the event horizon of the Vaidya-Bonner black hole,
namely,
\begin{equation}
r_{H}=\frac{M+\sqrt{M^2-Q^2(1-2\dot{r}_H)}}{1-2\dot{r}_{H}}=(M-\omega)+\sqrt{M^2-Q^2}-\frac{M^2}{\sqrt{M^2-Q^2}}\frac{\omega}{M}+O(\frac{\omega}{M})^2.\label{horizon22}
\end{equation}
Since the tunnelling out point of the barrier in Parikh-Wilczek
tunnelling framework is
\begin{equation}
r_f=(M-\omega)+\sqrt{(M-\omega)^2-Q^2}=(M-\omega)+\sqrt{M^2-Q^2}-\frac{M^2}{\sqrt{M^2-Q^2}}\frac{\omega}{M}+O(\frac{\omega}{M})^2,\label{horizon23}
\end{equation}
we have
\begin{equation}
r_{H}=r_f=(M-\omega)+\sqrt{(M-\omega)^2-Q^2}.\label{horizon24}
\end{equation}
Similar to the dynamical Schwarzschild, the event horizon of the
dynamical Reissner-Nordstr\"{o}m black hole corresponds to the
turning point $r_f=(M-\omega)+\sqrt{(M-\omega)^2-Q^2}$, that is,
the event horizon $r_H$ is a turning point of the tunnelling
barrier, and the ergosphere region coincide with the barrier. The
ergospere region corresponds to the classical inhibited region.
\section{where does the Hawking radiation  come from?}
Let us move to the final question: where does the Hawking
radiation come from? According to the Parikh-Wilczek tunnelling
framework, the ingoing tunnelling of a virtual negative energy
particle is equivalent to the outgoing tunnelling of a real
positive energy particle across the barrier from $r_i(M-\omega)$
to $r_f(M)$. Since the region between the event horizon $r_H$ and
the apparent horizon $r_{AH}$ is the ergosphere, which is
classically not able to detect, the emitted particle can only be
detected outside the $r_{AH}$. In this meaning, we see that the
Hawking radiation comes from the apparent horizon.
\section{conclusion}
In above calculation, we treated the background space times in the
Parikh-Wilczek tunnelling framework as a non-stationary
counterparts. By using the Parikh-Wilczek tunnelling method the
positions of the event horizons of the Vaidya black hole and the
Vaidya-Bonner black holes were calculated. We find that the
apparent horizon and the event horizon of these two types of black
holes correspond to the two turning points of their tunnelling
barriers. The barrier region corresponds to the ergospere, and
therefore the ergosphere is a classical inhibited region. Since
the region between the event horizon and the apparent horizon is
classically inhibited, we think that the Hawking radiation comes
from the apparent horizon. \acknowledgments This research is
supported partly by the National Natural Science Foundation of
China (Grant No. 10873003), and the Natural Science Foundation of
Guangdong Province (Grant No. 7301224).

\end{document}